\documentstyle[seceq,epsf,wrapft]{ptptex}
\def\dt{{\delta \theta}}

\def\dt2{(\delta \theta)^2}

\def\p{\sigma}
\def\al{\alpha}

\def\ga{\gamma}

\def\lan{\left\langle}
\def\ran{\right\rangle}

\def\nonum{\nonumber}

\def\dt{\frac{\partial}{\partial T}}

\def\virg{,}
\def\point{.}
\def\e{{\it e }}
\def\vf{v_{\rm F}}
\def\kf{k_{\rm F}}
\def\lsim{\lower -0.3ex \hbox{$<$} \kern -0.75em \lower 0.7ex \hbox{$\sim$}}
\def\gsim{\lower -0.3ex \hbox{$>$} \kern -0.75em \lower 0.7ex \hbox{$\sim$}}
\def\yen{Y \kern -1.077em =}
\def\jo #1#2#3#4{#1 {\bf #2} (#3),  #4}   %For Prog. Theor. Phys.
\def\PR{Phys.\ Rev.}

\def\JDP{J.\ de\ Phys.}

\def\JPSJ{J.\ Phys.\ Soc.\ Jpn.}

\def\PTP{Prog.\ Theor.\ Phys.}

\def\ADV{Adv.\ Phys.}

\def\SL{JETP\ Lett.}

\def\PL{Phys.\ Lett.}

\def\IJMP{Int.\ J.\ Mod.\ Phys.}

\def\MCLC{Mol.\ Cryst.\ Liq.\ Cryst.}
\pubinfo{Vol. 98, No. 5, November 1997}  %Editorial Office use
\notypesetlogo  %comment in if to eliminate PTPTeX logo
\title{%        %You can use \\ for explicit line-break
Crossover between High and Low Energy-States \\ 
in Two-Coupled Chains of Tomonaga Model
}

\author{%       %Use \sc for the family name
 Masahisa {\sc Tsuchiizu}, Hideo {\sc Yoshioka} 
 and Yoshikazu {\sc Suzumura}\\ 
}

\inst{%         %Affiliation, neglected when [addenda] or [errata]
Department of Physics, Nagoya University,
Nagoya 464-01 
}
\recdate{%      %Editorial Office will fill in this.
 July 22, 1997}

\abst{%         
By applying the renormalization group method to  
 two-coupled chains in the Tomonaga model,  
 the role of interchain hopping 
 has been  studied in the entire energy region.  
 The energy for a crossover from   the perturbational regime to  
 the relevant regime 
  becomes smaller than that of 
 the interchain hopping  due to one-dimensional fluctuations 
 of the mutual interaction.    
   From the calculation of response  functions for 
 charge density  waves and superconducting states,
  the phase diagram of 
    dominant and subdominant states  has been obtained  
   in the plane of mutual interactions 
  with fixed energy.  
}
\begin{document}
\maketitle
%
%---------------------------------------------
\section{Introduction}
  Organic conductors,
 \cite{Ishiguro_Yamaji} 
  which exhibit  low dimensional fluctuations,  
 have been studied by using the model of   
 quasi-one-dimensional  systems,   
 where the role of the interchain hopping plays an important role.
 
  As a basic model for such a system, 
 two-coupled chains of interacting electrons has been explored. 
  Even in the simple case of spinless fermions described by 
  the Tomonaga model, there exist  significant  properties 
 involved in connecting a one-dimensional system 
  with  quasi-one dimensional system.
% \cite{Wen,Kusmartsev,Yakovenko,Nersesyan,Yoshioka_Suzumura_JPSJ94,Yoshioka_Suzumura_JPSJ95} 
 \cite{Wen}$^-$\cite{Yoshioka_Suzumura_JPSJ95} 
 The ground state has been calculated in the limit of low energy.
  The relevant behavior of the interchain hopping  leads to 
   a  state  which   differs  essentially from  that in the absence  of  hopping.  
 Although the limiting state for the two-coupled chains 
  of the Tomonaga model  is well known,  it is not yet clear 
 how the state at finite temperature (or energy) is determined 
  in the presence of both 
   the interchain hopping  and the mutual interactions.

 The phase diagram in the plane of the intrachain interaction 
  ($g_2$) and the interchain interaction ($g_2'$) is shown  
  in  Fig. 1(a) (Fig. 1(b)).
  \cite{Yoshioka_Suzumura_JPSJ95} 
 This represents the result  in the absence (presence) of  
 the interchain hopping. 
 In Fig. 1(a), there are two kinds of dominant states 
 with  in-phase pairing  and out-of-phase pairing
  which are degenerate due to the absence of interchain hoping. 
  Such a state corresponds  to the state in the high energy limit,
    where the effect of the interchain hopping can be disregarded.   
  The state in Fig. 1(b) is obtained in the limit of zero energy  
  (or temperature). 
  The interchain hopping plays a crucial role 
  in the sense that  the degeneracy for
    in-phase and out-of-phase pairings is removed 
 and  the other pairing state becomes  subdominant 
  due to the relevance of the interchain hopping. 
   The subdominant state which  is shown 
 in the parenthesis  could become meaningful for more 
 complicated models, e.g., a model with  a Fermi surface 
  which exhibits an  incomplete  nesting condition 
  in the case   $|g_2| > |g_2'|$.   
  Thus, it is of interest 
  to examine  the state  with finite energy, 
  which is expected to exhibit a phase diagram between  
 Figs. 1(a) and (b). 

%------------------(Fig.1)-------------------------------------------
\newpage
\begin{figure}[tb]
  \parbox{\halftext}{
     \epsfxsize = 10cm
     \vspace{-2.5cm}
     \centerline{\epsfbox{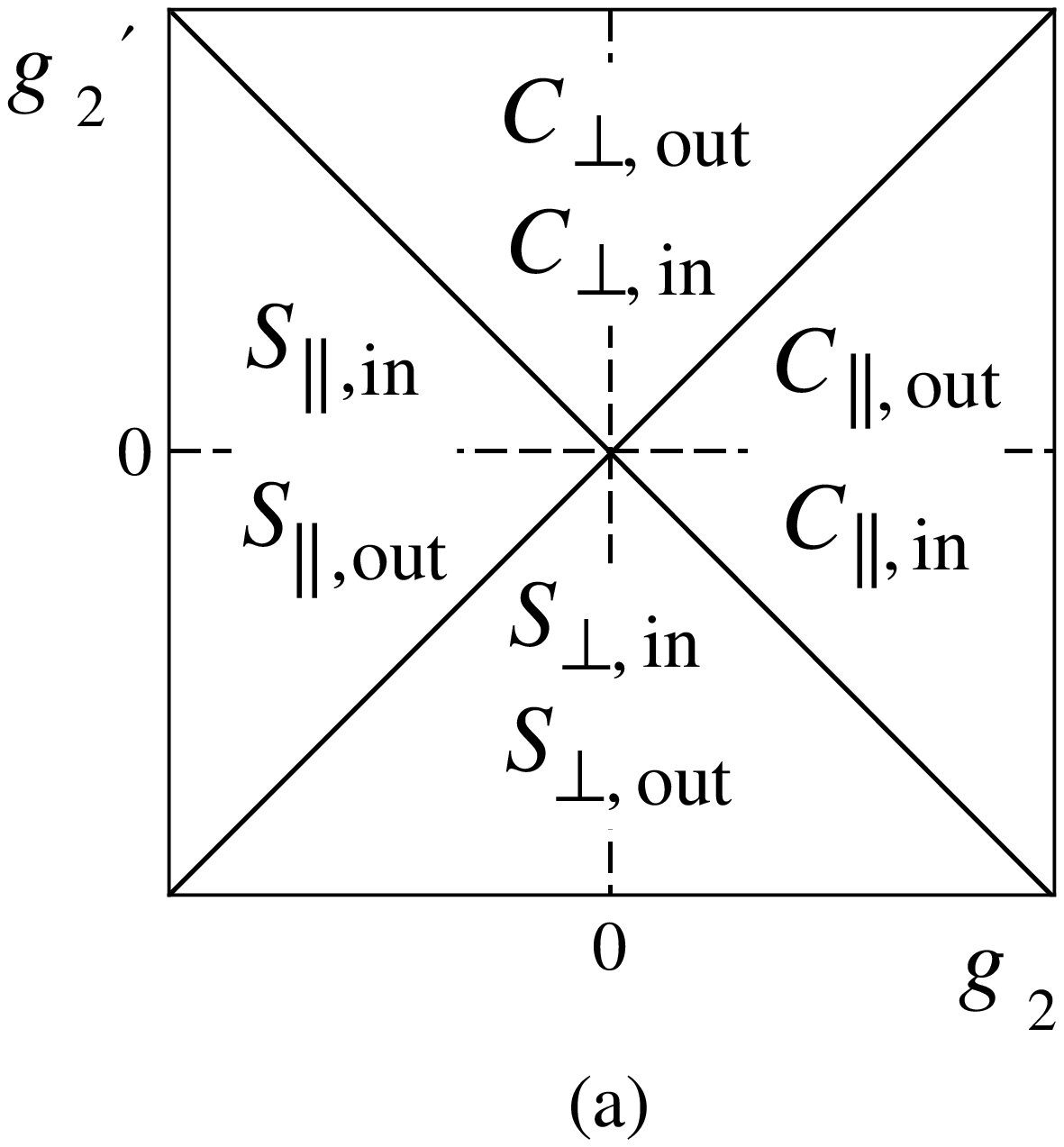}}
%     \vspace{-5.0cm}
  }
  \hspace{5mm}
  \parbox{\halftext}{
     \epsfxsize = 10cm
     \vspace{-2.5cm}
     \centerline{\epsfbox{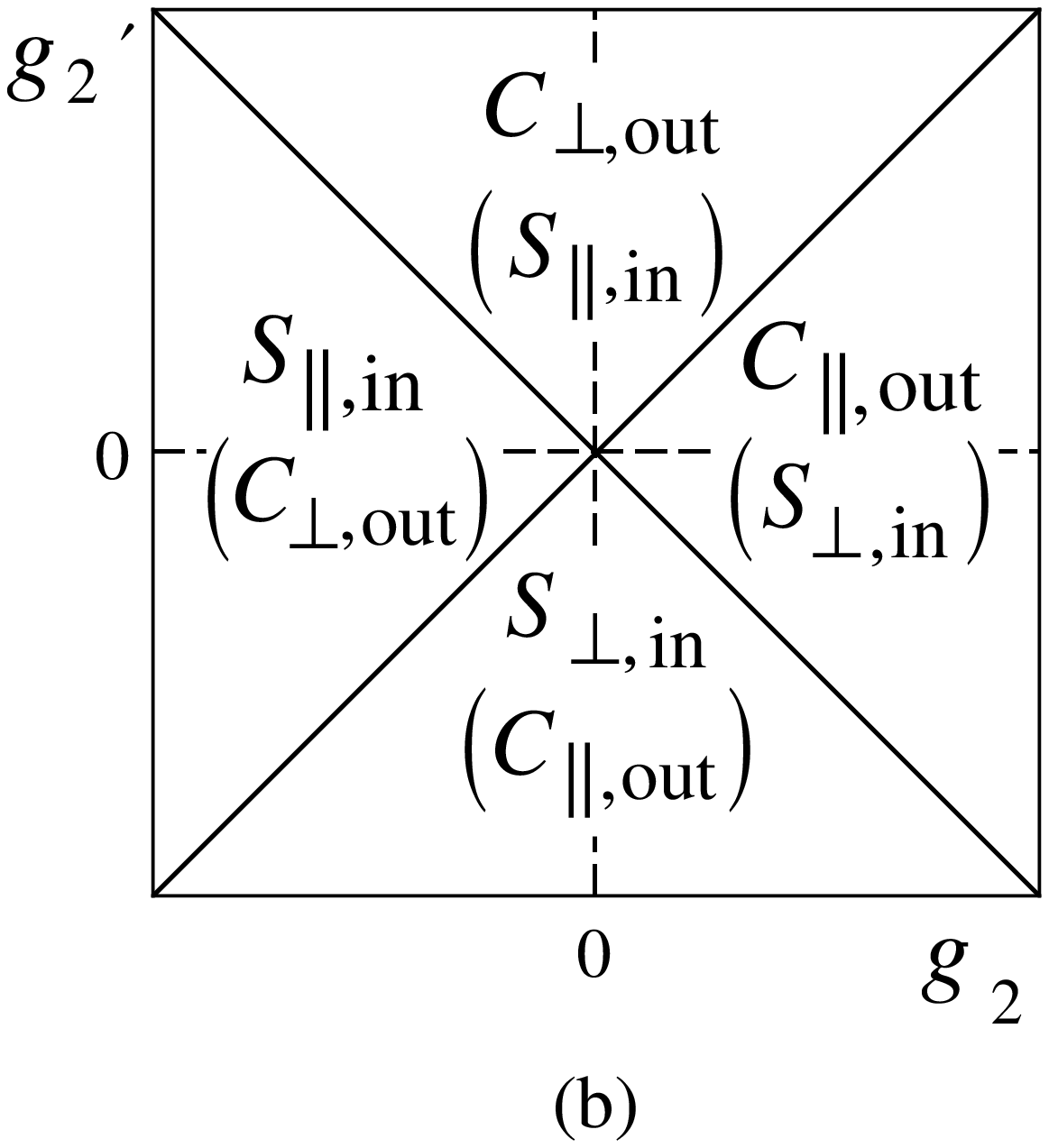}}
%     \vspace{-5.0cm}
  }
\vspace{-5.5cm}
\caption{
Phase diagram 
 in the plane of  $g_2=\gamma _2/(2\pi \vf)$ and 
 $g_2'=\gamma_2'/(2\pi \vf)$ for $t=0$ (a) and $t \not= 0$ (b), 
 \protect{\cite{Yoshioka_Suzumura_JPSJ95}} 
 where 
 $C$ and $S$ denote the CDW and SC state, respectively,  
 and $\parallel$ ($\perp$) and out (in) denote 
 the paring state for the intrachain (interchain)  
 and  out of phase (in phase) ordering. 
 For $t=0$ (a),  there are two kinds of  states  as the dominant states,
  while for $t \not= 0$ (b),  
 the subdominant state is distinguished, as is shown in the parenthesis.
}
\label{fig.1}

\end{figure}
\vspace*{-1cm}
%----------------------------------------------------------------------
 In deriving Fig. 1(b),  it has been assumed that 
 one-dimensional behavior vanishes  and  
 the  relevant behavior of the interchain hopping begins  
 when the energy (or temperature) becomes smaller 
 than the  hopping energy. 
 In a quasi-one dimensional system, Bourbonnais
 \cite{Bourbonnais_MCLC85} 
  has  maintained  
 that one-dimensional fluctuations still exist at temperatures 
 lower than the hopping energy and that 
 the temperature for the crossover  
 becomes  smaller than the hopping energy 
 due to the suppression of the density of state around the Fermi energy.
 \cite{Suzumura_PTP80} 
   There is no explicit calculation which shows 
     the reduction of the crossover    temperature,   
    even for two-coupled chains.

In the present paper, 
 we study these problems in two-coupled chains 
 of the Tomonaga model by applying the renormalization group 
 method to the bosonized Hamiltonian.  
 In \S 2, the Hamiltonian and possible  order parameters are represented  in terms of the phase variables. 
   The scaling equations for  coupling constants and
  response functions   are  given in \S3. 
 In \S4, a phase diagram with a fixed 
  energy is examined, and  the crossover energy is evaluated 
   as a function of mutual interactions. 
   Section 5 is devoted to summary and discussion.

%-------------------------------------------------------
\section{Model and order parameter}
We consider  two-coupled chains of the Tomonaga model, 
 where a one-dimensional chain consists  of  spinless fermions with 
 intrachain forward scattering, interchain forward scattering, 
 and  interchain hopping. 
The Hamiltonian with a chain of length $L$ is given as  
%---------- (1) -----------------------
\begin{eqnarray}
{\cal H}
&=& \sum _{p,i} \int dx \; \psi _{p,i}^\dagger (x) 
v_{\rm F} \left( -ip \partial_x - k_{\rm F} \right) \psi _{p,i}(x)
\nonumber \\
& & {} - t \sum _p \int dx \left\{ 
\psi_{p,1}^\dagger (x) \psi_{p,2}(x) + {\rm h.c.} \right\}
\nonumber \\
& & {} + \frac{\ga_2}{2} \sum_{p,i} \int dx \;
\psi_{p,i}^\dagger (x) \psi_{-p,i}^\dagger (x) \psi_{-p,i} (x) \psi_{p,i} (x)
\nonumber \\
& & {} + \frac{\ga'_2}{2} \sum_{p} \int dx 
\left\{
\psi_{p,1}^\dagger (x) \psi_{-p,2}^\dagger (x) \psi_{-p,2} (x) \psi_{p,1} (x)
+ {\rm h.c.} 
\right\} \virg
\label{eqn:Hreal}
\end{eqnarray}
%--------------------------------------
where $\psi_{p,i}^\dagger$ is the fermion  creation operator. 
The sign  $p=+$ $(-)$ expresses the right (left) moving state, 
 and $i=1$ $(2)$ denotes  the indices of the chain,
 where  $\vf$ and $\kf$ are the Fermi velocity 
and the Fermi wave vector, respectively. 
 We consider the case in which  the energy of interchain hopping, $t$, 
 is much smaller than the Fermi energy, $\vf \kf$, and 
 the matrix element of the intrachain (interchain) forward scattering, 
 $\ga_2 $ ($\ga'_2 $), is smaller than $2 \pi \vf$.  
  
In Eq. (\ref{eqn:Hreal}) the first and second terms 
  can be diagonalized 
  by use of the transformation
%---------- (2) -----------------------
\begin{equation}\psi_{p,\p} (x) = ( -\p \psi_{p,1} (x) + \psi_{p,2} (x) ) / \sqrt{2}
 \point
 \label{eqn:band}
\end{equation}
%--------------------------------------
 The index $\p = \pm$ denotes that for the diagonalized band, 
  where the new Fermi wave vector  is given by 
 $k_{{\rm F}\p} = \kf - \p t/\vf$ for the $\sigma$-band. 
 By applying the bosonization method to the respective band,  
 Eq. (\ref{eqn:Hreal}) is rewritten  
 in terms of the  phase Hamiltonian as   
${\cal H} = {\cal H}_\theta + {\cal H}_\phi$, 
 \cite{Yoshioka_Suzumura_JPSJ94,Yoshioka_Suzumura_JPSJ95}
 where  
%---------- (3) (4) -------------------
\begin{eqnarray}
{\cal H}_\theta &=& \frac{u}{4\pi}
\int dx \left\{ \frac{1}{K_\theta }(\partial _x \theta _+)^2 
+ K_\theta (\partial _x \theta _-)^2 \right\} 
\virg
\label{eqn:Htheta} \\
{\cal H}_\phi &=& \frac{v_{\rm F}}{4\pi}
\int dx \left\{ \frac{1}{K_\phi }(\partial _x \phi _+)^2 
+ K_\phi (\partial _x \phi _-)^2 \right\} 
\nonumber \\
& & {}+\frac{v_{\rm F}}{2 \pi \alpha ^2} \int dx 
\left\{ g_{2\phi +} \cos \left( 2\phi _+ - \frac{4t}{v_{\rm F}}x \right) 
+ g_{2\phi -} \cos \left( 2\phi _- \right) \right\}
\point
\label{eqn:Hphi}
\end{eqnarray}
%-----------------------------------
 Equations (\ref{eqn:Htheta}) and (\ref{eqn:Hphi}) 
  express the Hamiltonian of 
  the total fluctuation and that of the transverse fluctuation, 
  respectively, 
  where $u = \vf \sqrt{ 1-(g_2+g_2')^2 }$,  
$K_\theta
=\sqrt{ \left\{1-(g_2 + g_2')\right\}/\left\{1+(g_2 + g_2')\right\} }$, 
$K_\phi = 1$, $g_{2\phi +} = -(g_2 - g_2')$, 
$g_{2\phi -} = g_2 - g_2'$ and 
 $\ga_2 = 2 \pi \vf g_2$ ($\ga'_2 = 2 \pi \vf g_2'$). 
The quantity  $\al^{-1}$, which is of the order of $k_{\rm F}$,
  is the cutoff  for  large wave vectors. 
 The phase variables $\theta_\pm$ and $\phi_\pm$ 
  in Eqs. (\ref{eqn:Htheta}) and (\ref{eqn:Hphi})
  are defined as
%------------ (5) (6) ---------
\begin{eqnarray}
\theta _\pm (x) &=& - 
\sum_{q\ne 0} \frac{\pi i}{qL} \e^{-\frac{\alpha }{2}|q|+iqx} 
\sum_{k,\sigma} \left( a^\dagger_{k,+,\sigma }a_{k+q,+,\sigma } \pm 
a^\dagger_{k,-,\sigma }a_{k+q,-,\sigma } \right) \virg 
\label{eqn:theta}
\\
\phi _\pm (x) &=& - \sum_{q\ne 0} \frac{\pi i}{qL} \e^{-\frac{\alpha }{2}|q|+iqx} 
\sum_{k,\sigma} \sigma \left( a^\dagger_{k,+,\sigma }a_{k+q,+,\sigma } \pm 
a^\dagger_{k,-,\sigma }a_{k+q,-,\sigma } \right) \virg
\label{eqn:phi}
\end{eqnarray}
%-------------------------
where $a_{k,p,\p} = ( 1/\sqrt{L} ) \int dx \, \e^{-i k x} \, \psi_{p,\p}(x)$.  
 By use of Eqs. (\ref{eqn:theta}) and (\ref{eqn:phi}), 
 the field operator $\psi_{p,\p}(x)$ is expressed as
\cite{Luther_Peschel} 
%----------- (7) --------
\begin{equation}
\psi _{p,\sigma }(x) = \frac{1}{\sqrt{2 \pi \alpha }} 
\exp \left[ ipk_{{\rm F}\sigma }x + p\Theta_{p,\sigma } + i \pi \Xi _{p,\sigma } \right]
= \psi _{p,\sigma }'(x) \exp [i \pi \Xi_{p,\sigma}]
\virg
\label{eqn:field}
\end{equation}
%------------------------
where 
$\Theta_{p,\sigma } = i [\theta _+ +p\theta _- + \sigma (\phi_+ + p\phi_- )]/2$. 
The phase factor $i \pi \Xi_{p,\p}$, which is introduced 
 for the fermion operator, 
is given by $\Xi _{+,+} = 0$, $\Xi _{+,-} = \hat{N}_{+,+}$,  
$\Xi _{-,+} = \hat{N}_{+,+} + \hat{N}_{+,-}$ and 
$\Xi _{-,-} = \hat{N}_{+,+} + \hat{N}_{+,-} + \hat{N}_{-,+}$,
 where  $\hat{N}_{p,\sigma} = \int dx \, \psi_{p,\sigma}^\dagger (x) \,
 \psi_{p,\sigma}(x) $ is the number operator for the fermion. 

 We examine the state with order parameters for the 
 charge density wave (CDW) state and the superconducting (SC) state.
In terms of phase variables, order parameters corresponding to Fig. 1(a) 
  are expressed as follows.
%------------- (8) - (15) ----------
\begin{itemize}
%--------- CDW ------------
%------------- (8) -----------------
\item[(I)] Order parameters for CDW with intrachain and out of (in) phase pairing are given by
\begin{eqnarray}
\left(
   \begin{array}{c}
      C_{||,{\rm out}} \\
      C_{||,{\rm in}}
   \end{array}
\right)
&=& 
\left(
\begin{array}{c}
 \psi_{p,1}^\dagger \psi_{-p,1} - \psi_{p,2}^\dagger \psi_{-p,2} \\
 \psi_{p,1}^\dagger \psi_{-p,1} + \psi_{p,2}^\dagger \psi_{-p,2} 
\end{array}
\right) 
= 
\left(
\begin{array}{c}
 - {\displaystyle \sum_\sigma } \psi_{p,\sigma}^\dagger \psi_{-p,-\sigma} \\
   {\displaystyle \sum_\sigma } \psi_{p,\sigma}^\dagger \psi_{-p,\sigma} \\
\end{array}
\right)
\nonumber \\
&\rightarrow& 
\left(
   \begin{array}{c}
      {\displaystyle \sum_\sigma } \sigma {\psi '}_{p,\sigma}^\dagger 
                                           \psi '_{-p,-\sigma} \\
      {\displaystyle \sum_\sigma } {\psi '}_{p,\sigma}^\dagger 
                                    \psi '_{-p,\sigma}
   \end{array}
\right)
\nonumber \\
&=& \frac{\e^{-i2pk_{\rm F} x}}{\pi \alpha } \e^{-ip\theta _+} 
  \left(
     \begin{array}{c}
        - i \sin \phi _- \\
        \cos \left( \phi _+ - {\displaystyle \frac{2t}{v_{\rm F}} }x \right) 
     \end{array}
  \right)
\point
\label{eqn:Cpara}
\end{eqnarray}
%
%-------------- (9) --------------------------
\item[(II)] Order parameters  for CDW with interchain and out of (in) phase pairing are given by 
\begin{eqnarray}
\left(
   \begin{array}{c}
      C_{\bot,{\rm out}} \\
      C_{\bot,{\rm in}}
   \end{array}
\right)
&=& 
\left(
\begin{array}{c}
\psi_{p,1}^\dagger \psi_{-p,2} - \psi_{p,2}^\dagger \psi_{-p,1}  \\
\psi_{p,1}^\dagger \psi_{-p,2} + \psi_{p,2}^\dagger \psi_{-p,1} 
\end{array}
\right)
= 
\left(
\begin{array}{c}
  -{\displaystyle \sum_\sigma }\sigma \psi_{p,\sigma}^\dagger 
                                      \psi_{-p,-\sigma}  \\
  -{\displaystyle \sum_\sigma }\sigma \psi_{p,\sigma}^\dagger 
                                      \psi_{-p,\sigma} 
\end{array}
\right)
\nonumber \\
&\rightarrow& 
\left(
   \begin{array}{c}
      {\displaystyle \sum_\sigma }{\psi '}_{p,\sigma}^\dagger 
                                    \psi '_{-p,-\sigma} \\
      {\displaystyle \sum_\sigma } \sigma {\psi '}_{p,\sigma}^\dagger 
                                            \psi '_{-p,\sigma}
   \end{array}
\right)
\nonumber \\
&=& \frac{\e^{-i2pk_{\rm F} x}}{\pi \alpha } \e^{-ip\theta _+} 
  \left(
     \begin{array}{c}
        \cos \phi _- \\
       -ip\sin \left( \phi _+ -{\displaystyle \frac{2t}{v_{\rm F}}}x \right) 
     \end{array}
  \right)
\point 
\label{eqn:Cperp}
\end{eqnarray}

%--- super -----
%---------- (10) ----------------------------
\item[(III)] Order parameters for the superconducting state  with 
intrachain  and in (out of) phase pairing are given by 
\begin{eqnarray}
\left(
   \begin{array}{c}
      S_{||,{\rm in}} \\
      S_{||,{\rm out}}
   \end{array}
\right)
&=&
\left(
\begin{array}{c}
\psi_{p,1} \psi_{-p,1} + \psi_{p,2} \psi_{-p,2}  \\
\psi_{p,1} \psi_{-p,1} - \psi_{p,2} \psi_{-p,2} 
\end{array}
\right) 
=
\left(
\begin{array}{c}
  {\displaystyle \sum_\sigma } \psi_{p,\sigma} \psi_{-p,\sigma}  \\
 -{\displaystyle \sum_\sigma } \psi_{p,\sigma} \psi_{-p,-\sigma} 
\end{array}\right) 
\nonumber \\
&\rightarrow& 
\left(
   \begin{array}{c}
      {\displaystyle \sum_\sigma } \psi '_{p,\sigma} \psi '_{-p,\sigma} \\
      {\displaystyle\sum_\sigma }\sigma \psi '_{p,\sigma}\psi '_{-p,-\sigma}
   \end{array}
\right)
\nonumber \\
&=& \frac{1}{\pi \alpha } \e^{-ip\theta _-} 
  \left(
     \begin{array}{c}
        \cos \phi _- \\
        ip\sin \left( \phi _+ -{\displaystyle \frac{2t}{v_{\rm F}}}x \right) 
     \end{array}
  \right)
\point 
\label{eqn:Spara}
\end{eqnarray}
%
%----------- (11) ------------------------------
\item[(IV)] Order parameters for the superconducting state with 
interchain and in (out of) phase pairing are given by
\begin{eqnarray}
\left(
   \begin{array}{c}
      S_{\bot ,{\rm in}} \\
      S_{\bot ,{\rm out}}
   \end{array}
\right)
&=& 
\left(
\begin{array}{c}
\psi_{p,1} \psi_{-p,2} + \psi_{p,2} \psi_{-p,1}  \\
\psi_{p,1} \psi_{-p,2} - \psi_{p,2} \psi_{-p,1} 
\end{array}
\right)
= 
\left(
\begin{array}{c}
 - {\displaystyle \sum_\sigma } \sigma \psi_{p,\sigma} \psi_{-p,\sigma}  \\
 - {\displaystyle \sum_\sigma } \sigma \psi_{p,\sigma} \psi_{-p,-\sigma} 
\end{array}
\right)
\nonumber \\
&\rightarrow& 
\left(
   \begin{array}{c}
      {\displaystyle \sum_\sigma }\sigma \psi '_{p,\sigma} 
                                         \psi '_{-p,\sigma} \\
      {\displaystyle \sum_\sigma }\psi '_{p,\sigma} \psi '_{-p,-\sigma}
   \end{array}
\right)
\nonumber \\
&=& \frac{1}{\pi \alpha } \e^{-ip\theta _-} 
  \left(
     \begin{array}{c}
        i \sin \phi _- \\
        \cos \left( \phi _+ - {\displaystyle \frac{2t}{v_{\rm F}}} x \right) 
     \end{array}
  \right)
\point
\label{eqn:Sperp}
\end{eqnarray}
\end{itemize}
%-----------------------------------
 The symbol  $\parallel$ ($\perp$) represents
 the pairing in the chain (between  two chains), and 
 the symbol in (out) denotes  pairing with the same  (opposite) sign. 
The right arrow in the second line of Eqs. (\ref{eqn:Cpara})$\sim$(\ref{eqn:Sperp})
  denotes the process obtained from Eq. (\ref{eqn:field}), where
  a numerical factor  
  with  absolute value of unity is discarded.     
In Eq. (\ref{eqn:Hphi}) and 
 Eqs. (\ref{eqn:Cpara})$\sim$(\ref{eqn:Sperp}), 
 the quantity $N_{+,+} - N_{-,+}$ is taken as an even integer, where
$N_{p,\p}$ is an eigenvalue of $\hat N_{p,\p}$. 
We note that the symmetry of the interactions is associated 
  with the four kinds of order parameters of 
  Eqs. (\ref{eqn:Cpara})$\sim$(\ref{eqn:Sperp}).
Since a quarter of the $g_2$-$g_2'$ plane in Fig. 1(a) is sufficient to
  understand all the states, we precisely examine  the states 
  in the region  $g_2>|g_2'|$.
 
%-------------------------------------------------
%=========    III ============
\section{Renormalization group method} 
Several kinds of order parameters introduced in the previous section 
  are examined by 
  calculating the response functions given by 
%----------- (12) ----------------
\begin{eqnarray} 
R_A (x,\tau)
&=&
\lan T_\tau A ^\dagger (x,\tau) A (0,0) \ran \cdot 2(\pi \al )^2
\nonumber \\
&=&
R_A ^\theta (x,\tau) \cdot R_A ^\phi (x,\tau) 
\virg
\label{eqn:response}
\end{eqnarray}
%---------------------------------
%---------------------------------
 where $A$ is the operator given by Eqs. (\ref{eqn:Cpara})$\sim$(\ref{eqn:Sperp}).
 From Eq. (\ref{eqn:Htheta}), 
 $R_A^{\theta}(x,\tau)$  is calculated straightforwardly  as 
%----------- (13) ----- ----------
\begin{equation}
R_A^\theta (x,\tau)
=
\lan T_\tau \e^{i\theta _\pm (x,\tau)} \e^{-i\theta _\pm (0,0)} \ran 
=
\left( \frac{\alpha }{\sqrt{x^2 + (\vf \tau)^2}} \right)^{K_\theta ^{\pm 1}}
\virg
\label{eqn:col-theta}
\end{equation}
%-----------------------------------------
  where the sign $+$ ($-$)  corresponds to the CDW (SC) state. 
Note that Eq. (\ref{eqn:col-theta}) depends only on
  $r(=\sqrt{x^2+(\vf \tau)^2})$.
From Eqs. (\ref{eqn:Cpara})$\sim$(\ref{eqn:Sperp}), 
  the quantity  $R_A^{\phi}(x,\tau)$
  corresponding to  the $\phi$-field is expressed as   
%-----------  (14) ----------------
\begin{eqnarray} 
 R_A ^\phi (x,\tau) 
 = 
 \overline{R_A ^\phi} (x,\tau) \cdot f_A(x)
\virg
\label{eqn:res_phi}
\end{eqnarray}
%--------------------------------- 
 where 
%------------ (15) --------------
\begin{equation}
\overline{R_A ^\phi} (x,\tau)
=
\left\{
 \begin{array}{l}
 R_{\sin \phi _-} \equiv 2\lan T_\tau \sin \phi_-(x,\tau)\sin \phi_-(0,0)\ran
     \\ \hspace{160pt} 
     \mbox{ for $A=C_{\parallel ,{\rm out}},S_{\perp ,{\rm in}}$ }, \\
 R_{\cos \phi _-} \equiv 2\lan T_\tau \cos \phi_-(x,\tau)\cos \phi_-(0,0)\ran
     \\ \hspace{160pt}
     \mbox{ for $A=C_{\perp ,{\rm out}},S_{\parallel ,{\rm in}}$ }, \\
 R_{\sin \phi _+} \equiv \displaystyle{
           \frac{ 2 \lan T_\tau \sin \left( \phi_+(x,\tau) 
                     + 2tx/\vf \right) \sin \phi_+(0,0) \ran }
                { \cos \left( 2tx/\vf \right) } }
     \\ \hspace{160pt}
     \mbox{ for $A=C_{\perp ,{\rm in}},S_{\parallel ,{\rm out}}$ }, \\
 R_{\cos \phi _+} \equiv \displaystyle{
           \frac{ 2 \lan T_\tau \cos \left( \phi_+(x,\tau) 
                     + 2tx/\vf \right) \cos \phi_+(0,0) \ran }
                { \cos \left( 2tx/\vf \right) } }
     \\ \hspace{160pt}
     \mbox{ for $A=C_{\parallel ,{\rm in}},S_{\perp ,{\rm out}}$ }
 \end{array}
\right.
    \label{eqn:res_b}
\end{equation}
%--------------------------------
 and the factor  $f_A(x)$ is defined as  
$f_A(x) \equiv 
  e^{i2pk_{\rm F}x}$   for $A = C_{\parallel ,{\rm out}}$ 
                                  and $C_{\perp ,{\rm out}}$, 
$f_A(x) \equiv$ 
  $\e^{i2pk_{\rm F}x} \cos \left( 2tx/\vf \right)$ 
  for $A = C_{\parallel ,{\rm in}}$   and $C_{\perp ,{\rm in}}$ , 
$f_A(x) \equiv   1$ 
 for $A = S_{\parallel ,{\rm in}}$  and $S_{\perp ,{\rm in}}$,  and 
 $f_A(x) \equiv \cos \left( 2tx/\vf \right)$ 
for $A = S_{\parallel ,{\rm out}}$   and $S_{\perp ,{\rm out}}$.  

By using the renormalization group method
  with the assumption of invariance with respect to 
  $\al \to \al'=\al \e^{dl}$,
  response functions for the  $\phi_\pm$-field
  are derived as (Appendix A)
%--------------- (16) - (19) ----------
\begin{eqnarray}
R_{\sin \phi _-} (r)
&=& 
2 \left< T_\tau \sin \phi _-(x,\tau) \sin \phi _-(0,0) \right> 
\nonumber \\
&=& \exp \left[ -\int_0^{\ln (r/\alpha )}\frac{1}{K_\phi (l)} dl \right]
\cdot \exp \left[ +\int_0^{\ln (r/\alpha )}g_{2\phi -}(l) dl \right]
\virg
\label{eqn:Rsin-} \\
%-----------------------
R_{\cos \phi _-} (r)
&=& 
2 \left< T_\tau \cos \phi _-(x,\tau) \cos \phi _-(0,0) \right> 
\nonumber \\ 
&=& \exp \left[ -\int_0^{\ln (r/\alpha )}\frac{1}{K_\phi (l)} dl \right]
\cdot \exp \left[ -\int_0^{\ln (r/\alpha )}g_{2\phi -}(l) dl \right] 
\virg
\label{eqn:Rcos-} \\
%-----------------------
R_{\sin \phi _+} (r)
&=& 
2 \left< T_\tau \sin \left( \phi _+(x,\tau) -\frac{2t}{v_{\rm F}} x \right) 
\sin \phi _+(0,0) \right> \bigg/ \cos \left( \frac{2t}{v_{\rm F}} x \right)
\nonumber \\
&=& \exp \left[ -\int_0^{\ln (r/\alpha )} K_\phi (l) dl \right] \cdot 
\exp \left[ +\int_0^{\ln (r/\alpha )}g_{2\phi +}(l) dl \right]
\virg
\label{eqn:Rsin+} \\
%----------------------
R_{\cos \phi _+} (r)
&=& 
2 \left< T_\tau \cos \left( \phi _+(x,\tau) -\frac{2t}{v_{\rm F}} x \right) \cos \phi _+(0,0) \right> \bigg/ \cos \left( \frac{2t}{v_{\rm F}} x \right) 
\nonumber \\
&=& \exp \left[ -\int_0^{\ln (r/\alpha )} K_\phi (l) dl \right] \cdot 
\exp \left[ -\int_0^{\ln (r/\alpha )}g_{2\phi +}(l) dl \right]
\virg
\label{eqn:Rcos+}
\end{eqnarray}
%------------------------------------------------------
 where the term with the quantity 
 $\tan ^{-1} (\vf \tau /x)$ has been neglected
 for the present numerical calculation with small $t$. 
 In Eqs. (\ref{eqn:Rsin-})$\sim$(\ref{eqn:Rcos+}), 
 the quantities $K_{\phi}(l)$, $g_{2\phi -}(l)$ and $g_{2\phi +}(l)$ 
 are calculated from the renormalization  equations, 
%-------- (20) - (23) --------  
\begin{eqnarray}
\frac{d}{dl} K_\phi(l) &=& 
-\frac{1}{2}g_{2\phi +}^2(l) \, K_\phi^2(l) \, J_0\left( \frac{4t(l)}{\vf \al ^{-1}} \right) 
+ \frac{1}{2}g_{2\phi -}^2(l)
\virg 
\label{eqn:rg-K} \\
%---------------------
\frac{d}{dl} g_{2\phi +}(l) &=& \left( 2-2K_\phi(l) \right) \, g_{2\phi +}(l) 
\virg
\label{eqn:rg-gp} \\
%---------------------
\frac{d}{dl}  g_{2\phi -}(l) &=& \left( 2-2/K_\phi(l) \right) \, g_{2\phi -}(l) 
\virg
\label{eqn:rg-gm} \\
%---------------------
\frac{d}{dl} \left( \frac{4t(l)}{\vf \al ^{-1}} \right) &=& 
\frac{4t(l)}{\vf \al ^{-1}} 
- g_{2\phi +}^2(l) \, K_\phi(l) \, J_1\left( \frac{4t(l)}{\vf \al ^{-1}} \right) 
\virg
\label{eqn:rg-t} 
\end{eqnarray}
%---------------------
where 
$J_\nu(z)$ is the $\nu$-th Bessel function. 
Equations (\ref{eqn:Rsin-})$\sim$(\ref{eqn:rg-t}) have been derived 
  by use of a method similar to that of Giamarchi and Schulz. 
  \cite{Giamarchi_JPF88,Giamarchi_PRB89}
The initial conditions in Eqs. (\ref{eqn:rg-K})$\sim$(\ref{eqn:rg-t})
  are chosen as 
 $K_\phi(0)=1$, $g_{2\phi +}(0)=-(g_2-g_2')$, 
  $g_{2\phi -}(0)=g_2-g_2'$ and 
 $t(0)=t$. 
We note that 
  the renormalization equation (\ref{eqn:rg-t})
  for  small $t$ becomes equal to  that derived 
  from the perturbation in terms of $t$.
\cite{Nersesyan} 
The negative sign of the second term of  Eq. (\ref{eqn:rg-t}) 
  \cite{Yoshioka_err}
 indicates the reduction of  interchain hopping.

 From Eqs. (\ref{eqn:col-theta}), 
 (\ref{eqn:Rsin-})$\sim$(\ref{eqn:Rcos+}), 
 the  response function can be expressed as 
%---------- (24) -------------------------------
\begin{eqnarray}
            \label{eqn:res_AB}
\overline{R_{A}}(r) = R_{A}^{\theta}(x,\tau) 
   \overline{R_{A}^{\phi}}(x,\tau) \virg 
\end{eqnarray}
%-----------------------------------------------
 where the largest response function and the next one 
 in  $\overline{R_{A}}(r)$ 
 with  fixed length $r (= \sqrt{ x^2 + (\vf \tau)^2})$ 
 (i.e., the corresponding energy $\vf/r$)  
 lead to the phase diagram 
 for the dominant state  and subdominant state, respectively.

 In the present paper, 
  we assume that  the response function $R_A(r)$ 
   is equivalent to that of the Fourier  transform 
    with $\omega = \vf/r$. 
  The validity of such a treatment is discussed in \S5.

%=========================================
\section{Response functions and phase diagram} 

%-----------------(Fig.2)----------------------------------------------
\begin{wrapfigure}{r}{6.6cm}
 \epsfysize=7cm
 \centerline{\epsfbox{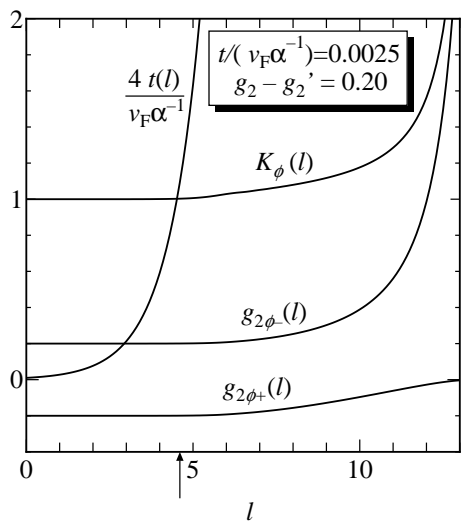}}
\caption{
 Quantities $K_{\phi(l)}$, $g_{2\phi-}(l)$, $g_{2\phi+}(l)$ 
and $4t(l)/(\vf \al^{-1})$ 
as functions of $l = \ln (r/\al )$  
for $t/(v_{\rm F}\alpha ^{-1}) = 0.0025$ and $g_2-g_2' = 0.20$. 
The arrow denotes $l = l_0 = \ln [\vf \al^{-1}/4t] \, (\simeq 4.6)$.
}
\label{fig.2}
\end{wrapfigure}
%----------------------------------------------------------------------
We examine  the renormalization group equations given by 
 Eqs. (\ref{eqn:rg-K})$\sim$
 \\
 (\ref{eqn:rg-t}),
 where parameters are chosen as 
  $t/(\vf \al ^{-1}) = 0.0025$ and $g_2 - g_2' = 0.02$. 
 The $l$-dependence  of 
$ K_{\phi}(l)$, $g_{2\phi-}(l)$ and $g_{2\phi+}(l)$ are shown 
 in Fig. 2. 
Increasing $l$, 
  $K_{\phi}(l)$ and $g_{2\phi-}(l)$ increases to the strong coupling regime,  while  $g_{2\phi+}(l)$ reduces to zero. 
 Note that the quantities $d(l)$  of Eq. (\ref{eqn:re-d}), 
 which is obtained by substituting 
 the results of Eqs. (\ref{eqn:rg-K})$\sim$(\ref{eqn:rg-t}), 
 is invisible in the scale of Fig. 2.  
 The actual variation of $g_{2\phi-}(l)$ and $g_{2\phi+}(l)$ 
 appears for $l \gsim \ln [\vf \al ^{-1} /4t ] \sim 4.6$, 
  which is designated by the arrow.  
  This result indicates the validity of 
 the  previous treatment
\cite{Yoshioka_Suzumura_JPSJ95} 
 that  
 $K_\phi$ and $g_{2\phi \pm}$ have been  replaced by 
 the initial values   
 for $e^{l} < \vf \alpha^{-1}/t $.  
 Since  the Bessel function $J_0 (4t(l)/(\vf \al^{-1}))$ 
  becomes small  
 for $e^{l}  > \vf \alpha^{-1}/t $,   
 the $\phi_+$-term can be 
%-----------------(Fig.3)----------------------------------------------
\newpage
\begin{wrapfigure}{r}{6.6cm}
 \epsfysize=6.8cm
 \centerline{\epsfbox{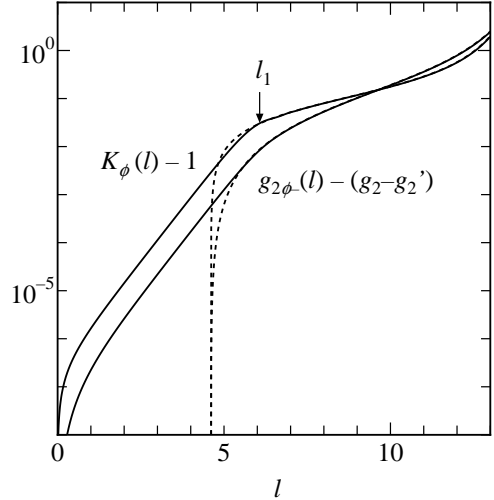}}
\caption{
 The $l$-dependence of  $K_{\phi}(l) -1$ and 
 $g_{2\phi-}(l)-(g_2-g_2')$,  which denote the variation  
 from the initial values.  
 The parameters are the same as in Fig. 2. 
 The dotted curves represent  the previous results,
\protect{\cite{Yoshioka_Suzumura_JPSJ94}} 
 which were obtained 
 by use of the initial condition 
 that 
 $g_{2\phi+}(l_0) = 0$, 
$K(l_0) = 1$ and $g_{2\phi -}(l_0) = g_2-g_2'$
with $l_0 = \ln [v_{\rm F}\alpha ^{-1}/4t]$.
 The factor 2 error in the renormalization equation 
 of the previous papers
\protect{\cite{Yoshioka_Suzumura_JPSJ94,Yoshioka_Suzumura_JPSJ95}}
 is corrected.
The quantity $l_1$ denotes $l$ for the onset for the relevant behavior 
 of $K_{\phi}(l)$ in the presence of  $t$. 
}
\label{fig.3}
\end{wrapfigure}
\noindent
%----------------------------------------------------------------------
neglected and leads to  strong coupling 
 due to the  $\phi_-$-term.  
%------------------------------
 Equation (\ref{eqn:rg-t}) reveals  
  the fact that 
 the hopping energy is reduced by  the mutual interaction.  
 Therefore  the range for the one-dimensional regime 
is examined in detail.

%-------------   Fig.3  ------------
In Fig. 3 
the present results (solid curves)  are compared with 
 the previous  ones (dotted curves), which were obtained by 
 using  initial conditions, such that 
 the terms including $t(l)$ are discarded 
  in Eqs. (\ref{eqn:rg-K}) and (\ref{eqn:rg-t})
 and for which  $K(l_0) = 1$ and 
$g_{2\phi -}(l_0) = g_2 - g_2'$
with $l_0 = \ln [\vf \al ^{-1}/4t]$.
\cite{Yoshioka_Suzumura_JPSJ94,Yoshioka_Suzumura_JPSJ95} 
In  the solid curve of $K_{\phi}(l)-1$, 
  the tangent, which displays the linear dependence 
   for  $ 1 \lsim l \lsim 5$,
  exhibits a rapid variation for  $l \simeq 6$, 
  indicating the crossover of the scaling property from 
  the one-dimensional regime to the regime of relevant  hopping.    
Therefore we define such a point, $l=l_1$, by the condition 
  that the absolute value of the variation of the tangent becomes maximum. 
%---------------------------
Good agreement between the solid curve and the dashed curve 
 is obtained for $l > l_1$, 
 where $l_1 > \ln[ \vf \alpha^{-1}/4t ]$. 
 For the small $l$ ($\ll l_1$),  Eqs. (\ref{eqn:rg-K})$\sim$(\ref{eqn:rg-gm})
 up to the lowest order of  $t/(\vf \al^{-1})$ and $g_2 - g_2'$
 are  calculated as (Appendix B)
%------------ (25) - (27) -------------
\begin{eqnarray}
K_\phi (l) &=& 1 + (g_2 - g_2')^2 
\left( \frac{t}{\vf \al ^{-1}} \right)^2 
\left( \e^{2l} - 1 \right)
+ \cdots  \virg
\label{eqn:K-app-1} \\
%---------------
g_{2\phi +} (l) &=& - (g_2 - g_2') + (g_2 - g_2')^3 
\left( \frac{t}{\vf \al ^{-1}} \right)^2 
\left( \e^{2l} - 1 - 2l \right) 
+ \cdots  \virg
\label{eqn:gp-app-1} \\
%------------------
g_{2\phi -} (l) &=&  (g_2 - g_2') + (g_2 - g_2')^3 
\left( \frac{t}{\vf \al ^{-1}} \right)^2 
\left( \e^{2l} - 1 - 2l \right)
+ \cdots  \point
\label{eqn:gm-app-1}  
\end{eqnarray}
%----------------------------------------

Equations (\ref{eqn:K-app-1}) and (\ref{eqn:gm-app-1}) 
 can reproduce the numerical results sufficiently,
 i.e., the linear dependence of the solid curves
 for $l< l_1$ in Fig. 3. 
Thus it turns out that there is a perturbational effect of $t$ 
 for the one-dimensional regime which is 
 obtained for  $l < l_1$.  
 The dependence of the crossover energy on the interaction 
 is discussed later.

%----------(Fig.4)-----------------------------------------------------
\begin{wrapfigure}{r}{6.6cm}
 \epsfysize=7cm
 \centerline{\epsfbox{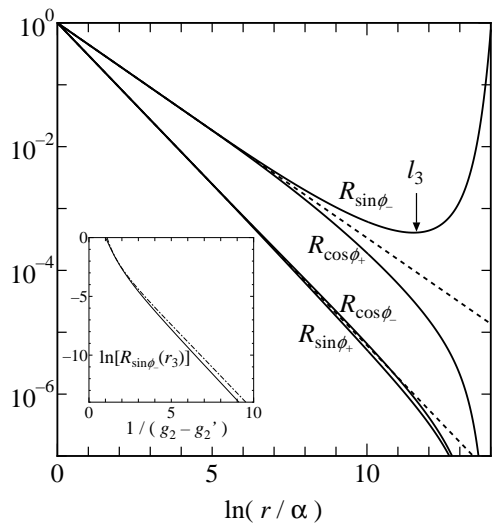}}
\caption{
 The response functions of 
 the $\phi_\pm$-field as a function of $\ln (r/\al)$,
 where $r=\protect\sqrt{ x^2 + (\vf \tau)^2 }$. 
 The quantities   $R_{\sin \phi_-}(r) , R_{\cos \phi_-}(r), 
  R_{\sin \phi_+}(r)$ and $R_{\cos \phi_+}(r)$ 
are defined by  
 Eqs. (\protect{\ref{eqn:Rsin-}}), 
 (\protect{\ref{eqn:Rcos-}}), (\protect{\ref{eqn:Rsin+}}) 
 and  (\protect{\ref{eqn:Rcos+}}), respectively. 
 The parameters are the same as in Fig. 2.
The dotted lines denote the  response functions with $t=0$, and  
  the location, $l_3(=\ln(r_3/\al))$, corresponds to a minimum of 
 $R_{\sin \phi_-}(r)$. 
 In the inset,  $\ln [R_{\sin \phi_-}(r_3)]$ (solid curve)
 as a function of $g_2-g_2'$  is compared with 
  the analytical one (the dash-dotted curve).  
}
\label{fig.4}
\end{wrapfigure}
%----------------------------------------------------------------------
%----------- Fig.4  ------------
 By substituting  the solutions of 
 Eqs. (\ref{eqn:rg-K})$\sim$(\ref{eqn:rg-t}) into 
 Eqs. (\ref{eqn:Rsin-})$\sim$(\ref{eqn:Rcos+}), 
 we obtain response functions for $\phi_{\pm}$ 
 which are shown by the solid curve  in Fig. 4 
  for $t/(\vf \al^{-1})=0.0025$ and $g_2 - g_2'=0.2$. 
 The dotted line corresponds to those  in the absence of the hopping. 
 For $\ln (r/\al) <l_1 (\simeq  6)$,  
the response functions of $R_{\sin \phi_-}(r)$ and $R_{\cos \phi_+}(r)$ 
($R_{\cos \phi_-}(r)$ and $R_{\sin \phi_+}(r)$) are almost the same as
 that for $t=0$ (dashed line), 
 and their deviations from the dashed line 
 due to the interchain hopping become visible 
 for $\ln (r/\al) > l_1$.    
 Since the renormalization group method is based 
 on  perturbation  with respect to 
$g_{2\phi+}$ and $g_{2\phi-}$, 
 the present result is valid only 
  for the regime of  weak  coupling.  
For  large $\ln (r/\al)$, 
 $R_{\sin \phi_-}(r)$ increases due to the strong coupling
 but is actually reduced to a finite value.     
 We define $l_3(\equiv \ln (r_3/\al))$ as 
the location for the crossover 
 from the weak coupling regime to the strong coupling regime  
 which corresponds to a minimum of $R_{\sin \phi_-}(r)$. 
As is shown in Appendix C,
 the excitation gap, $\Delta$,  in Eq. (\ref{eqn:Hphi}) 
 can be estimated from 
  $\Delta = \vf \al^{-1} \exp [-l_\Delta ]$
  with $ 2(K_\phi (l_\Delta ) -1) \simeq 1$, i.e.,
%------------ (28) -----------------
\begin{equation}
\Delta =4t \exp \left[ -\frac{\pi}{2} \frac{1}{|g_2-g_2'|} + 1 \right]
                                            \point
\label{eqn:gap-analytical}
\end{equation}
%-----------------------------------
In the inset in Fig. 4, $\ln R_{\sin \phi_-}(r_3)$ (solid curve)
  as a function of $1/(g_2 - g_2')$ 
  is compared with the  analytical function $R_{\sin \phi_-}(r_\Delta )$
  (dash-dotted curve) where $l_\Delta \equiv \ln (r_\Delta /\al)$. 
Since $r_3 \simeq r_\Delta$, it is found that the minimum of 
  $R_{\sin \phi_-}(r)$ at $r_3$ corresponds to the formation of the gap.

%------------ Fig. 5 -------------------
Now we examine 
 the actual response function of Eq. (\ref{eqn:res_AB}),
 which is a product of the response function of the total fluctuation 
  and that of the transverse one. 
 In Fig. 5,
  the response functions with 
  the largest (the dominant one), the  second largest 
  (the subdominant one) 
 and the third largest in   magnitudes  are shown 
  for $t/\vf \alpha^{-1}=0.0025$, $g_2 - g_2' = 0.20$ 
 and $g_2 + g_2' = 0.10$. 
In previous paper,
\cite{Yoshioka_Suzumura_JPSJ94,Yoshioka_Suzumura_JPSJ95} 
 the  dominant and the subdominant states were 
  examined in the limits of both  long range (Fig. 1(b)) 
and  $t \rightarrow 0$.
 In the present study, 
 by tracking a state with  arbitrary length scale 
 (corresponding to the inverse of the energy) 
 we found that the crossover of the second dominant states 
 from $C_{\parallel,{\rm in}}$ to $S_{\perp,{\rm in}}$ 
 occurs at   $\ln (r/\al) = l_2$. 
In the region of $\ln (r/\al) <l_1$,
 there is still a small effect of  the interchain hopping 
 which removes the degeneracy of $C_{\perp,{\rm out}}$ 
  and $C_{\parallel,{\rm in}}$. 
Such an effect can be seen analytically: 
 Eq. (\ref{eqn:res_AB}) with $\ln (r/\al) \ll l_1$  is expanded as
%-------------(Fig.5)-(Fig.6)-------------------------------------------
\begin{figure}[tb]
 \parbox{\halftext}
{
 \epsfysize=7cm
 \centerline{\epsfbox{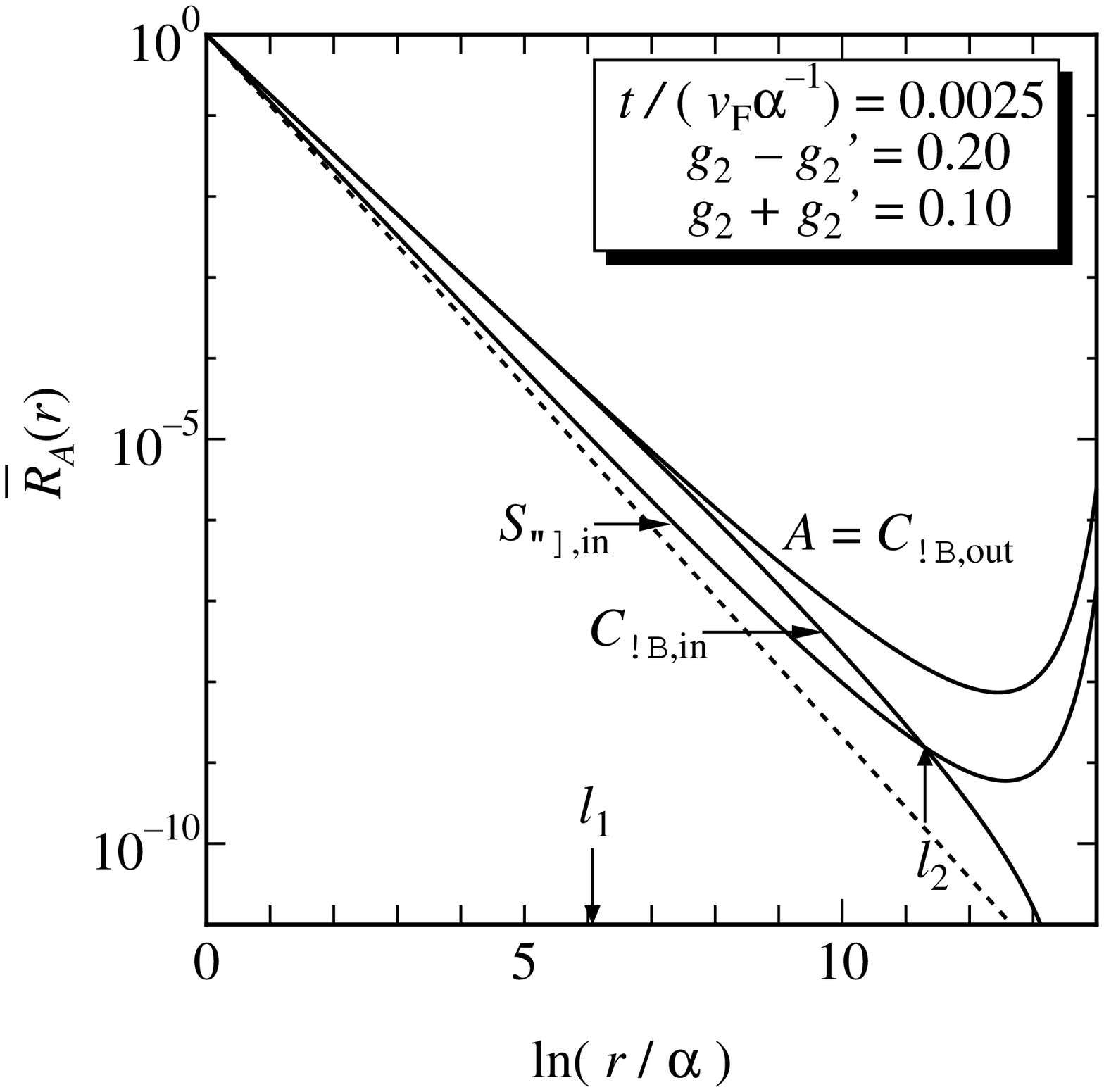}}
\caption{
The normalized response functions, 
  $\overline{R_{A}}(r)$,
  of the order parameters, 
$C_{\parallel,{\rm out}}$, $C_{\parallel,{\rm in}}$ and $S_{\perp,{\rm in}}$ 
  as functions of $\ln (r/\al)$, where 
 $t/(v_{\rm F}\alpha ^{-1}) = 0.0025$, $g_2-g_2' = 0.20$, 
  $g_2+g_2' = 0.10$ 
 and notation is the same as in Fig. 1. 
The dotted line denotes the response function in the absence of interactions.
 With increasing $\ln (r/\al)$, 
 the difference between $C_{\parallel,{\rm out}}$ 
 and $C_{\parallel,{\rm in}}$ becomes notable for $\ln (r/\al) > l_1$,
 and a subdominant response function varies from 
 $C_{\parallel,{\rm in}}$  to $S_{\perp,{\rm in}}$ 
 at   $\ln (r/\al) = l_2$.
}
\label{fig.5}
}
\hspace{8mm}
\parbox{\halftext}
{
 \epsfysize=6.8cm
 \centerline{\epsfbox{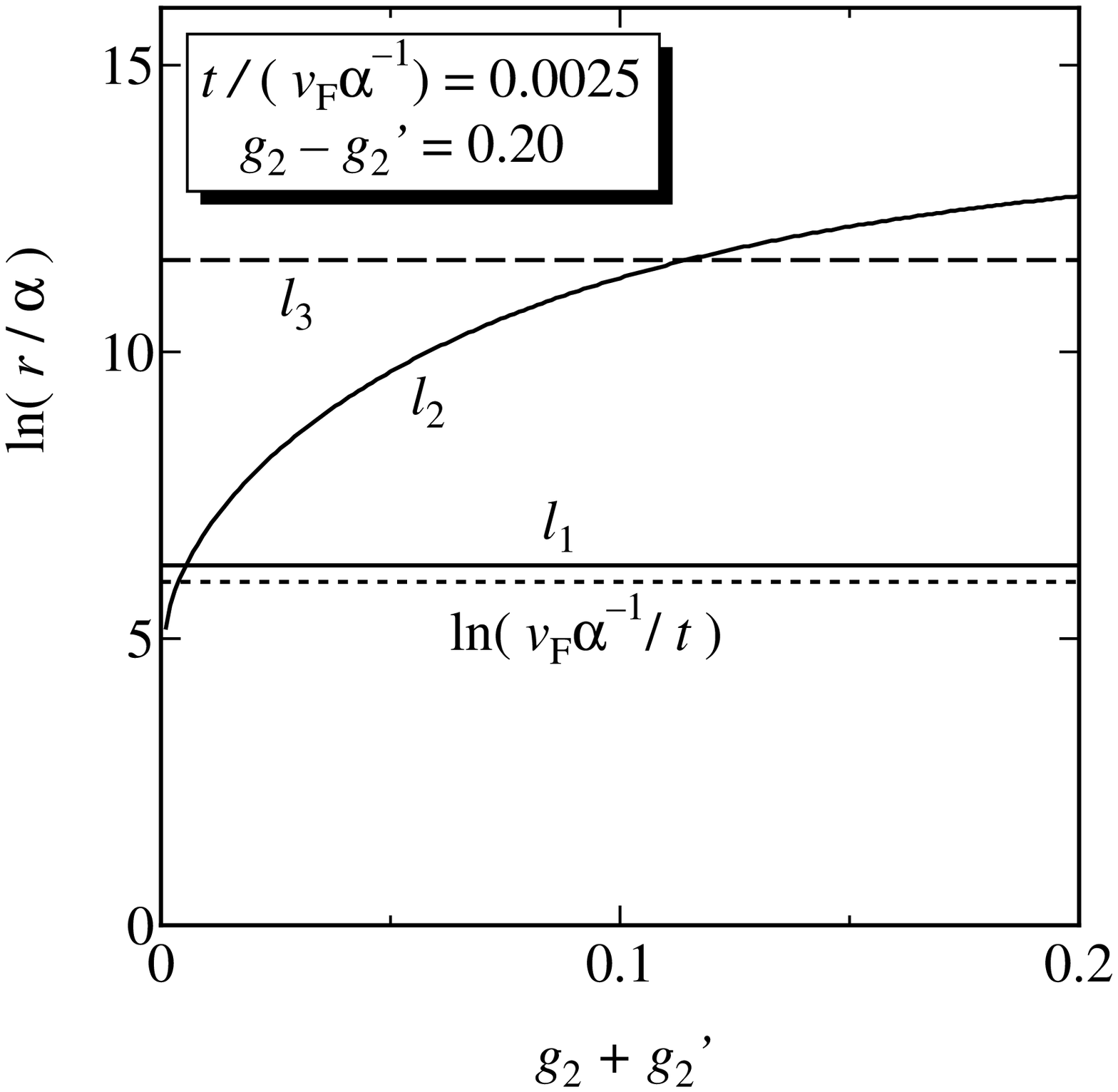}}
 \vspace{-1mm}
\caption{
Quantities $l_1$ and $l_2$ as  functions of $g_2+g_2'$,
 with fixed  $t/(v_{\rm F}\alpha ^{-1}) = 0.0025$ and 
 $g_2-g_2' =0.20$,  
where the one-dimensional regime, i.e., the irrelevant $t(l)$ 
 is obtained for    $0 < \ln (r/\al) < l_1$ and 
  the subdominant state is given by 
 $C_{\parallel,{\rm in}}$ ($S_{\perp,{\rm in}}$)  for $\ln (r/\al)<l_2$ 
  ($\ln (r/\al)>l_2$). 
  The value $l_3$ (dashed line) is the upper bound 
 for the present treatment of the renormalization group method. 
 The maximum value for the one-dimensional regime denoted by $l_1$ 
  is larger than  $\ln [\vf \al^{-1}/4t]$ (dotted line).  
}
\label{fig.6}
}
\end{figure}
%----------------------------------------------------------------------
%-------------- (29)-(31) ------------
\begin{eqnarray}
\overline{R_{C_{\parallel ,{\rm out}}}} 
    && 
(r)
  =  \left( \frac{\al }{r} \right) ^{K_\theta }
            \left( \frac{\al}{r} \right) ^{1-(g_2-g_2')}
                                         \nonumber \\
\times && 
         \left[ 1 + \frac{(g_2-g_2')^2}{32} 
            \left( \frac{4t(0)}{\vf \al ^{-1}}\right) ^2
                    \left\{ \e^{2\ln (r/\al)}-1-2\ln \left( 
                    \frac{r}{\al}\right) \right\} + \cdots \right]
                                    \virg 
         \label{eqn:CDW-out-app} \\
%-------------------------
\overline{R_{C_{\parallel ,{\rm in}}}} 
    && 
(r)
        = \left( \frac{\al }{r} \right) ^{K_\theta }
          \left( \frac{\al}{r} \right) ^{1-(g_2-g_2')}
                               \nonumber \\
\times && 
      \left[ 1-\frac{(g_2-g_2')^2}{32} 
\left( \frac{4t(0)}{\vf \al ^{-1}}\right) ^2
\left\{ \e^{2\ln (r/\al)}-1-2\ln \left( \frac{r}{\al}\right) \right\} 
+ \cdots \right]
                                  \virg 
   \label{eqn:CDW-in-app} \\
%--------------------------
\overline{R_{S_{\perp,{\rm in}}}} 
    && 
(r)
= \left( \frac{\al }{r} \right) ^{1/K_\theta }
\left( \frac{\al}{r} \right) ^{1-(g_2-g_2')}
\nonumber \\
\times && 
      \left[ 1+\frac{(g_2-g_2')^2}{32} 
\left( \frac{4t(0)}{\vf \al ^{-1}}\right) ^2
\left\{ \e^{2\ln (r/\al)}-1-2\ln \left( \frac{r}{\al}\right) \right\}
+ \cdots \right]
\point
\label{eqn:SC-in-app}
\end{eqnarray}
%---------------------------------------------------------   
For the range satisfying  $\ln (r/\al) > l_3$
  corresponding to an energy lower than $\Delta$,
  there appears a gap in the transverse fluctuation 
  due to the relevance of the interchain hopping which leads to  
  a large enhancement  for $C_{\parallel ,{\rm out}}$ 
  and $S_{\perp ,{\rm out}}$ and  strong  
  suppression for  $C_{\parallel ,{\rm in}}$. 
We have found two kinds of quantities, $l_1$ and $l_2$, 
 where coupling constants are renormalized toward 
  the relevant $t$ in the region   $\ln (r/\al)>l_1$  and 
  $S_{\perp,{\rm in}}$ is found as
  the subdominant state
due to  the relevant $t$  in the region   $\ln (r/\al)>l_2$. 
 In Fig. 6, $l_1$ and $l_2$ are shown as a function of $g_2+g_2'$. 
 The dashed line denotes $l_3$ corresponding to the upper bound 
 of $\ln (r/\al)$ for the  present treatment of the renormalization group method 
 and the dotted line denotes $\ln [\vf \alpha^{-1}/t]$.     
In the region with $\ln (r/\al) <l_1$, the state exhibits  
 one-dimensional properties 
 since the  interchain hopping can be treated perturbatively.   
We note that $l_2 < l_1$ for small $g_2+g_2'$  due to 
  $l_2 \simeq (g_2+g_2')(g_2-g_2)^{-2}(\vf \al^{-1}/t)^2$
  which is obtained from Eqs. (\ref{eqn:CDW-out-app}), 
  (\ref{eqn:SC-in-app}) and $(g_2+g_2')l \ll 1$.

%----------(Fig.7)-----------------------------------------------------
\begin{wrapfigure}{r}{6.6cm}
 \vspace{-5mm}
 \epsfysize=7cm
 \centerline{\epsfbox{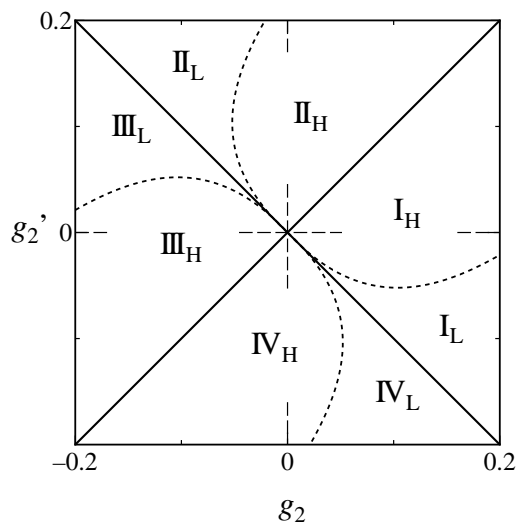}}
\caption{
 Phase diagram for several kinds of order parameters in the plane of 
 $g_2$ and $g_2'$, where  
 the dominant states (subdominant states) are given by 
 $C_{\parallel,{\rm out}}$ ($S_{\perp,{\rm in}})$, 
                 $C_{\parallel,{\rm out}}$ ($C_{\parallel,{\rm in}}$), 
 $C_{\perp,{\rm out}}$ ($S_{\parallel,{\rm in}})$, 
                 $C_{\perp,{\rm out}}$ ($C_{\perp,{\rm in}}$), 
 $S_{\parallel,{\rm in}}$ ($C_{\perp,{\rm out}})$, 
                 $S_{\parallel,{\rm in}}$ ($S_{\parallel,{\rm out}}$) and  
 $S_{\perp,{\rm in}}$ ($C_{\parallel,{\rm out}})$, 
                 $S_{\perp,{\rm in}}$ ($S_{\perp,{\rm out}}$) for 
 the regions  
 ${\rm I}_{\rm L}$,  ${\rm I}_{\rm H}$,
 ${\rm II}_{\rm L}$,  ${\rm II}_{\rm H}$,
 ${\rm III}_{\rm L}$,  ${\rm III}_{\rm H}$ and 
 ${\rm IV}_{\rm L}$,  ${\rm IV}_{\rm H}$,  respectively. 
 The dotted  curve, which corresponds to the boundary for the 
 subdominant states 
 is calculated for  $\ln(r/\al)=\ln (\vf \al^{-1}/ \omega) = 12$. 
}
\label{fig.7}
\end{wrapfigure}
%----------------------------------------------------------------------
Based on the results of Figs. 5 and 6, we obtain 
 the phase diagram as a function of  $g_2$ and $g_2'$   
 for the dominant and subdominant states 
   with  fixed $\ln (r/\alpha)$.  
 The boundary between  
  $C_{\parallel, {\rm in}}$ and $S_{\perp, {\rm in}}$ 
  is obtained from Fig. 6 by estimating $g_2$ and $g_2'$ 
  which satisfy  $\ln(r/\alpha) = l_2$ with  fixed $r$.
 For example,  the subdominant state, 
  $S_{\perp,{\rm in}}$, with $\ln(r/\alpha) = 12$  
  appears  in case  $g_2-g_2'=0.20$ and $g_2+g_2'< 0.11$
  due to the enhancement of the charge fluctuation.
%-------------- Fig.7 ----------------------
We note that the response function with  fixed $r$ 
 corresponds to the response function for  finite energy 
  with $\omega = \vf/r$.  
 In Fig. 7, the phase diagram    in the $g_2$-$g_2'$ plane is shown 
  with $t/(\vf \al^{-1})=0.0025$ and  fixed $\ln (r/\alpha) =12$.
The solid line denotes the boundary for the dominant state
  and the dotted line is  the boundary for the subdominant state.
%-------------------------------
The results of Figs. 5 and 6 correspond to the state in the region  
  $g_2>|g_2'|$, i.e., ${\rm I_L}$ and ${\rm I_H}$,
  which show the states with
  low energy and high energy, respectively.
The phase diagram of the entire region has been obtained by use of the 
  symmetry property of order parameters which is given by 
  Eqs. (\ref{eqn:Cpara})$\sim$(\ref{eqn:Sperp}).
The region of ${\rm J_L}$ (${\rm J=I}$, ${\rm II}$, ${\rm III}$ 
  and ${\rm IV}$) 
  which is similar to the state in the zero limit of energy
  decreases rapidly by the decrease of 
  $\ln (r/\al)$.
We have found that the region of ${\rm J_L}$ becomes very narrow 
  in case  $\ln (r/\al) \simeq l_1$
  ($\simeq \ln [\vf \al^{-1} /t] $).
Therefore  
  the phase diagram is similar to Fig. 1(a) for most of the region 
  in the $g_2$-$g_2'$ plane in the case that 
  $\ln (r/\al) <l_1$, 
  i.e., for energy larger than $t$.

%-----------(Fig.8)-----------------------------------------------------
\begin{wrapfigure}{r}{6.6cm}
 \epsfysize=7cm
 \centerline{\epsfbox{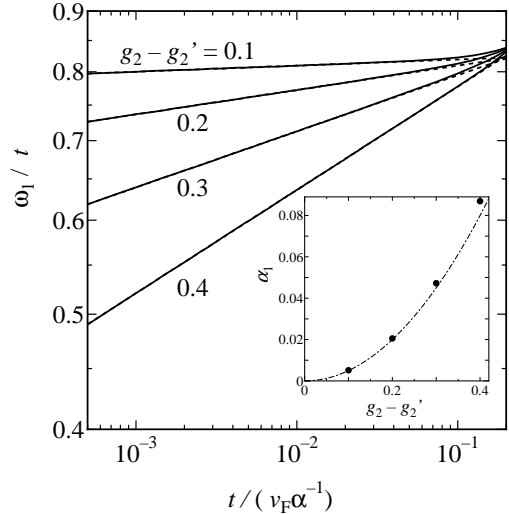}}
 \vspace{-3mm}
\caption{
The ratio of $\omega _1$ to $t$ 
 as a function of $t/(v_{\rm F} \alpha ^{-1})$,
where $\omega_1 \equiv \vf \alpha^{-1} \exp[-l_1]$.  
 The dotted line is  the  extrapolation 
  which is given by using the equation  
 $\omega_1/t  \propto  (t/\vf \alpha^{-1})^{\alpha_1}$.   
 In the inset, $\alpha_1$ is shown as a function of $g_2-g_2'$,
  where the dash-dotted curve denotes 
  $\alpha_1 = (1/2)(g_2-g_2')^2$.
}
\label{fig.8}
\end{wrapfigure}
%----------------------------------------------------------------------
Finally we examine the characteristic energy $\omega_1$ 
  which separates the one-dimensional regime 
  from the relevant regime of $t$.
The length corresponding to $\omega_1$ is shown by
  $l_1 = \ln [\vf \al^{-1}/\omega_1]$ in Fig. 3.
In Fig. 8 the quantity $\omega_1/t$ is shown 
  as a function of $t/(\vf \al^{-1})$ 
  with  fixed $g_2-g_2'=0.1$, $0.2$, $0.3$ and $0.4$. 
The quantity $\omega_1/t$ decreases by the increase of $g_2-g_2'$
  and increases with increasing $t/(\vf \al^{-1})$.
The dotted line which is obtained by use of the extrapolation 
  of the solid curve is expressed as 
  $\omega_1/t=\mbox{const} \cdot [t/(\vf \al^{-1})]^{\al_1}$,
  where $\al_1$ depends on $g_2-g_2'$.
Such a characteristic energy $\omega_1$ can be estimated 
  from Eq. (\ref{eqn:rg-t}) by assuming that 
  $\omega_1 \simeq \omega_1'=\vf \al^{-1} \exp [-l_1']$ with
  $t(l_1')/(\vf \al^{-1})=O(1)$,  where $l_1'\simeq l_1$.
Since Eq. (\ref{eqn:rg-t}) is calculated as 
  $t(l)/(\vf \al^{-1})=t/(\vf \al^{-1}) \exp [ \{ 1-(g_2-g_2')^2/2 \} l]$
  for $t/(\vf \al^{-1}) \ll 1$, we obtain 
%----------- (32) --------------------------
\begin{equation}
\omega_1' =
\mbox{const} \cdot t \left( t/\{\vf \al ^{-1}\} \right) ^{(g_2-g_2')^2 /2}
\point
\label{eqn:omega1}
\end{equation}
%------------------------------------------
In the inset of Fig. 8, $\al_1$ obtained from the data of the solid curve 
  is represented by the closed circle which is compared with the 
  $\al_1=(g_2-g_2')^2/2$ of Eq. (\ref{eqn:omega1}) (dash-dotted curve).
The good agreement of the two results indicates the validity of 
  Eq. (\ref{eqn:omega1}) as a characteristic energy $\omega_1$.
Here we note the assertion of Bourbonnais\cite{Bourbonnais_MCLC85}
  that  one-dimensional fluctuation does exist at  temperatures
  lower than the energy of the transverse hopping $t_\perp$
  in quasi-one-dimensional systems.
The suppression of the crossover temperature $T_x$ ($< t_\perp$)
  in a quasi-one-dimensional system
  comes from the reduction of the density of states at the Fermi energy,
  which is caused by the mutual interaction.
  \cite{Suzumura_PTP80}
We find that $\omega_1$ of the present calculation is essentially 
  the same as $T_x$ since a system of two-coupled chains already shares 
  common features with a quasi-one-dimensional system 
  regarding the role of interchain hopping.

%===========================================================================
\section{Summary and discussion}
In the present paper, 
 we have examined the role of  interchain hopping 
  in the entire energy region  
  in  two-coupled chains of the Tomonaga model
  by use of the renormalization group method. 
From the calculation of response functions for CDW  
  and SC states in the real space,
  we obtained the phase diagram in the plane of mutual interactions
  in which the dominant and subdominant states are shown 
  with fixed energy or temperature.
The interchain hopping exhibits a  perturbational effect 
  for energy higher than $\omega_1$ ($= \vf \al^{-1} \e^{-l_1}$),
  while   the  degeneracy of in-phase and out-of-phase parings 
   is removed     due to the relevant behavior of the hopping 
    for  energy lower than $\omega_1$. 
 The hopping leads to  the formation of the excitation gap 
  in the transverse fluctuation of the $\phi$-field 
  for energy lower than $\vf \al^{-1} \e^{-l_3} $.
The crossover energy, $\omega_1$, is lower than $t$ 
  due to the interaction which gives rise to 
  the reduction of the density of states at the Fermi surface.

We have examined the two-chain system 
  by use of the renormalization group method in  real space 
 in order to examine the states at finite energy or 
  at finite temperature.   
We comment on the relation between the 
   present result and that derived by 
  the scaling of the energy cutoff, i.e., in  Fourier space.  
%---------------------------- 
 Fabrizio \cite{Fabrizio_PRB93}
 has studied the same model 
by applying  the renormalization group method with the scaling of energy
    to Eq. (\ref{eqn:Hreal}), 
 where  {\it all}  interactions are calculated perturbatively.   
  In the present calculation, the two-chain system is transformed into 
  the phase Hamiltonian through the bosonization method, 
  where only the non-linear terms   $g_{2\phi -}$ and $g_{2\phi +}$ 
  are treated perturbatively.  
By replacing  $l(=\ln(\al '/\al))$ with $-\ln(\omega_0'/\omega_0)$,
  where $\omega_0 (\sim \vf /\alpha)$ is the band cutoff,
  it is found that the renormalization equations of the coupling constants   obtained by Fabrizio are the same as ours, 
  Eqs. (\ref{eqn:rg-K})$\sim$(\ref{eqn:rg-t}),
  up to the lowest order of the coupling constants 
  and the interchain hopping. 
But there is a difference in the terms of 
  higher order in the interchain hopping. 
Invariance of the renormalization equations
   with respect to  $t \rightarrow -t$ exists  
 in  the present result, Eq. (\ref{eqn:rg-t}), 
 but does  not for  his results. 
 It is not yet clear if such a discrepancy 
 comes from the present choice of the phase Hamiltonian
  or from some other source. 
%------------------------------------

Finally, we discuss the response functions at finite energy.
 Since,  to our knowledge, 
  there is no calculation for   response functions 
 of   two-coupled chains   in terms of the renormalization 
   group  method  with an energy cutoff, 
   we mention   the case of no misfit parameter, 
  i.e., a one-dimensional Fermi gas. 
   The response function for  the energy cutoff
    is calculated  by applying the renormalization group method 
     to the quantity given by  
   $\overline{R}_i(0,\omega)
          =\pi \vf (\partial R_i(0,\omega)/\partial \ln \omega)$,
\cite{Solyom} 
  where $R_i(k,\omega )$ is the Fourier transform of $R_i(r)$. 
 By comparing Eqs. (\ref{eqn:Rsin-})$\sim$(\ref{eqn:Rcos+})
  in the case  $t=0$
\cite{Giamarchi_PRB89}   
 with 
   $\overline{R}_i(0,\omega)$,
\cite{Solyom} 
 it turns out that  
     $ R_i(r)/(\al/r)^2$ with   $r/\alpha = \omega_0 /\omega$ 
     can be replaced by  
$ \overline{R}_i(0,\omega)$ 
 when the $l$-dependences of $K_{\phi}(l)$ and $g_{2\phi \pm}(l)$   are neglected. 
  In the present calculation, 
  such a replacement is justified  for $l \lsim 10$ 
   and is meaningful   qualitatively 
   for  larger $l$.  Thus, 
   the present results may correspond to  
    those  at finite energy or at finite   temperature.

%------------Acknowledgemnt--------
\section*{ Acknowledgements }
 This work was partially supported by  a Grant-in-Aid 
 for Scientific  Research  (09640429)
 from the Ministry of Education, 
  Science, Sports and Culture.

%====================================================
%-------------------- Appendix ------------------ 
\appendix
%------------- Appendix A --------------
\section{Derivation of Renormalization Equations}
 The response functions are calculated 
 by use of  the renormalization group method as follows.
\cite{Giamarchi_JPF88} 
  Treating the nonlinear terms in Eq. (\ref{eqn:Hphi}), i.e.,  
  $g_{2\phi +}$- and $g_{2\phi -}$-terms, 
 as  the perturbation,  
  the response function for $R_{\sin \phi -}(x,\tau)$ of  
  Eq. (\ref{eqn:Rsin-}) up to the second order
 is calculated as 
%-------- (A1) ------------------------
\begin{eqnarray}
 & & R_{\sin \phi _-} (x_1-x_2,\tau_1-\tau_2)
= 2\lan T_\tau \sin \phi _- (x_1,\tau_1) \sin \phi _-(x_2,\tau_2) \ran
\nonumber \\
&=&
\exp \left[-\frac{1}{K_\phi}U(r_1-r_2)\right]
\nonumber \\
&\times&
\Bigg[
1 + \frac{1}{4\pi} g_{2\phi -} \e^{(2/K_\phi)U(r_1-r_2)}
    \int \frac{d^2 r_3}{\al ^2} 
    \e^{-(2/K_\phi)U(r_1-r_3)} \e^{-(2/K_\phi)U(r_2-r_3)}
\nonumber \\
& & 
  - \frac{1}{2} g_{2\phi +}^2 \int \frac{dr}{\al}
    \left( \frac{r}{\al} \right)^{3-4K_\phi} 
    J_0 \left( \frac{4t}{\vf} r \right) U(r_1-r_2)
\nonumber \\
& &
  - \frac{1}{4} g_{2\phi +}^2 \int \frac{dr}{\al}
    \left( \frac{r}{\al} \right)^{3-4K_\phi} 
    J_2 \left( \frac{4t}{\vf} r \right) \cos 2 \theta_{r_1-r_2}
\nonumber \\
& &
  + \frac{1}{2} g_{2\phi -}^2 \frac{1}{K_\phi ^2} \int \frac{dr}{\al}
    \left( \frac{r}{\al} \right)^{3-4/K_\phi} U(r_1-r_2) + \cdots
\Bigg]  \virg
 \label{eqn:AA1}
\end{eqnarray}
%--------------------------------------
where $r=\sqrt{x^2 + (\vf \tau)^2}$ , $U(r)=\ln(r/\al)$
and $ \tan \theta _r = \vf \tau /x$. 
$J_\nu (z)$ is the $\nu$-th Bessel function.
By raising the expansion of r.h.s. of Eq. (\ref{eqn:AA1}) 
 to  the exponential
 and  using    $\al $   as the lower cutoff 
  of the integrations,   one obtains  
%-------- (A2) ---------------------------------
\begin{eqnarray}
& &
R_{\sin \phi -}(x_1-x_2,\tau_1-\tau_2)
\nonumber \\
& &= \exp 
\left[ 
-\frac{1}{K_\phi^{\rm eff}} U(r_1-r_2)-d^{\rm eff} \cos (2\theta _{r_1-r_2}) 
\right]
\nonumber \\
& & \hspace{10pt}
\times
\left[
1 + \frac{1}{4\pi} g_{2\phi -} \e^{(2/K_\phi)U(r_1-r_2)}
       \int _\al \frac{d^2 r_3}{\al ^2} 
       \e^{-(2/K_\phi)U(r_1-r_3)} \e^{-(2/K_\phi)U(r_2-r_3)} + \cdots
\right]
\virg
\nonumber \\
\label{eqn:AA2}
\end{eqnarray}
%-----------------------------------------------
where the effective quantities are given by
%-------- (A3) ---------------------------------
\begin{eqnarray}
K_\phi^{\rm eff} 
&=&
K_\phi - \frac{1}{2} g_{2\phi +}^2 K_\phi^2 \int _\al \frac{dr}{\al} 
\left( \frac{r}{\al}\right) ^{3-4K_\phi} J_0\left(\frac{4t}{\vf}r\right)
\nonumber \\
& & \hspace*{5cm}
+\frac{1}{2} g_{2\phi -}^2 \int _\al \frac{dr}{\al}
\left( \frac{r}{\al}\right) ^{3-4/K_\phi}
\virg
\label{eqn:AA3}
\\ 
%------ (A4) -------------------------------
d^{\rm eff} 
&=&
d+\frac{1}{4} g_{2\phi +}^2 K_\phi^2 \int _\al \frac{dr}{\al} 
\left( \frac{r}{\al}\right) ^{3-4K_\phi} J_2\left(\frac{4t}{\vf}r\right)
\point
\label{eqn:AA4}
\end{eqnarray}
%------------------
The parameter $d$ is introduced in the zeroth order, 
where $d=0$ at $l=0$.
By assuming the scaling relations 
that Eqs. (\ref{eqn:AA3}) and (\ref{eqn:AA4}) are invariant 
for $\al \to \al '= \al \e ^{dl}$,
the renormalization equations for the coupling constants are obtained as
%----------- (A5)-(A8) -----------------------------
\begin{eqnarray}
\frac{d}{dl} K_\phi(l) &=& 
-\frac{1}{2}g_{2\phi +}^2(l) \, K_\phi^2(l) \, 
J_0\left( \frac{4t(l)}{\vf \al ^{-1}} \right) 
+ \frac{1}{2}g_{2\phi -}^2(l) 
\virg
\label{eqn:AA7}
\\
\frac{d}{dl} g_{2\phi +}(l) 
&=& \left( 2-2K_\phi(l) \right) \, g_{2\phi +}(l) 
\virg
\label{eqn:AA8}
\\
\frac{d}{dl}  g_{2\phi -}(l) 
&=& \left( 2-2/K_\phi(l) \right) \, g_{2\phi -}(l) 
\virg
\label{eqn:AA9}
\\
\frac{d}{dl} d(l) 
&=& \frac{1}{4} g_{2\phi +}^2(l) \, K_\phi^2(l) \, 
J_2\left( \frac{4t(l)}{\vf \al ^{-1}}\right)
\virg
\label{eqn:re-d}
\end{eqnarray}
%-----------------------------------------------
where $\al (0) =\al$ in the Bessel functions.
The renormalization for $4t(l)/(\vf \al^{-1})$
  is calculated as follows.
The actual difference 
  of the density between the two bands of Eq. (\ref{eqn:band}),
 $\Delta n$, is given by
%----------- (A9) ---------------------- 
\begin{equation}
\Delta n = 
 - 2t \al /\vf 
+ \frac{T}{L} \al \int dx d\tau \left< \partial _x \phi_+ \right>
\virg
\label{eqn:AA11}
\end{equation}
%---------------------------------------
where 
  $\partial _x \phi _+ = \pi \Sigma _p (\psi _{p,+}^\dagger \psi _{p,+} 
  - \psi _{p,-}^\dagger \psi _{p,-} )$. 
The renormalization equation for 
  $t(l)$  is derived by  assuming  the  scaling relation for 
  Eq. (\ref{eqn:AA11}).
The second term on the r.h.s of 
  Eq. (\ref{eqn:AA11}) is rewritten as 
%---------- (A12) -----------------------
\begin{eqnarray}
\int dx \, d\tau &&
\lan \partial _x \phi_+ \ran
=
\frac{1}{Z} {\rm Tr} \left[ \int dx \, d\tau \,
\partial _x \phi_+ \e^{-\int d\tau {\cal H}} \right]
\nonumber \\
&&=
\frac{1}{Z} {\rm Tr} \left[ 
 \frac{\partial}{\partial \lambda} 
\exp \left\{ -\int d\tau \, {\cal H} 
  + \lambda \int dx \, d\tau \, \partial _x \phi_+ \right\} 
  _{\lambda = 0} \right]
\virg
\label{eqn:AA12}
\end{eqnarray}
%----------------------------------------
  where $Z={\rm Tr} \exp [ -\int d\tau {\cal H} ]$ .
By writing $\tilde{\phi}_+ = \phi_+ - (2\pi / \vf) K_\phi \lambda x$
  in Eq. (\ref{eqn:AA12}), Eq. (\ref{eqn:AA11}) is calculated as 
%----------- (A15) ----------------------
\begin{eqnarray}
\Delta n
&=& - \frac{2t}{\vf \al^{-1}} 
+ \frac{2}{\alpha } g_{2 \phi +} K_\phi \frac{T}{L} 
\int dx \, d\tau \,
\left< x \sin \left( 2 \phi _+ - \frac{4t}{v_{\rm F}} x \right) \right>
\nonumber \\
&=&
 - \frac{2t}{\vf \al^{-1}} 
+ \frac{1}{2} g_{2 \phi +}^2 K_\phi 
\int _\alpha ^\infty \frac{dr}{\alpha} 
\left( \frac{r}{\alpha} \right) ^{2-4K}  
J_1 \left( \frac{4t}{v_{\rm F}} r \right)
+ \cdots
\point
\label{eqn:AA16}
\end{eqnarray}
%---------------------------------------
The  infinitesimal transformation 
  $\alpha' =\e^{dl} \alpha$ for Eq. (\ref{eqn:AA16}) 
  leads to
%------------ (A17) ---------------------
\begin{equation}
\frac{d}{dl} \left( \frac{4t(l)}{\vf \al ^{-1}} \right) =
\frac{4t(l)}{\vf \al ^{-1}} 
- g_{2\phi +}^2(l) \, K_\phi(l) \, J_1\left( \frac{4t(l)}{\vf \al ^{-1}} \right) 
\point
\label{eqn:AA17}
\end{equation}
%---------------------------------------
Note that  the present  model is similar to but slightly different 
  from the model of  a single chain of fermions with spin 1/2. 
  \cite{Giamarchi_JPF88}
The hopping $t$  corresponds to the external magnetic field
  in the latter model,
  where  the renormalization equation for the magnetization 
  has been derived
  by assuming   that the external magnetic field 
  satisfies the  scaling relations. 

The response functions are calculated by writing  
  $R_i (x,\tau) = F_i (r) \cdot
  \exp[-K_{\phi}^{\pm 1} \ln$ \\ 
 $(r/\al) - d \cos 2\theta_r]$,
  where  $i=\sin \phi _-,\cos \phi_-,\sin \phi_+ $ and $\cos \phi_+$
  and the $\pm$ sign corresponds the response function
  for the $\phi_\pm$-field.
By assuming the scaling relation  
%--------- (A18) --------------------------------
$
F_i \left( r , \al(l) , K_\phi(l) , g_{2\phi \pm}(l) \right)
=
I_i \left( dl , K_\phi(l) , g_{2\phi \pm}(l) \right)
\cdot F_i ( r ,\al(l+dl), K_\phi(l+dl), 
            g_{2\phi \pm}(l$ \\ $+dl) )
$,
%-----------------------------------------------  
  the multiplicative factor $I_i$ 
  for  $F_{\sin \phi -}$  is obtained as 
%--------- (A20) --------------------------------
\begin{eqnarray}
& &
I_{\sin \phi -}(dl,K_{\phi},g_{2\phi \pm})
\nonumber \\
& &=
\exp \Bigg[
g_{2\phi -} dl 
- \frac{1}{2} g_{2\phi +}^2 J_0 \left( \frac{4t}{\vf \al^{-1}} \right) 
U(r_1-r_2) dl 
\nonumber \\
& & 
-\frac{1}{4} g_{2\phi+}^2 J_2 \left( \frac{4t}{\vf \al^{-1}} \right)
\cos 2\theta_{r_1-r_2} dl
+\frac{1}{2} g_{2\phi-}^2 \frac{1}{K_\phi ^2} U(r_1-r_2) dl 
\Bigg]
\point
\label{eqn:AA20}
\end{eqnarray}
%-----------------------------------------------
Thus the functions $F_i$ are expressed as
\cite{Giamarchi_PRB89} 
%---------- (A21) ----------------------------------
\begin{eqnarray}
F_i (r , K_\phi , g_{2\phi \pm} )
&=&
\exp \left[ \sum_{l=0}^{\ln (r/\al)} \ln [ I_i(dl,  
                   K_\phi (l),g_{2\phi \pm} (l) ) ] \right]
\point
\label{eqn:AA21}
\end{eqnarray}
%---------------------------------------------------
Equations (\ref{eqn:AA17}) (\ref{eqn:AA20}) and (\ref{eqn:AA21}) lead to
%----------- (A25) ------------------------------------
\begin{equation}
R_{\sin \phi -} (x,\tau)
=
\exp \left[ \int_0^{\ln (r/\alpha )} dl
   \left\{ - \frac{1}{K_\phi (l)} + g_{2\phi -}(l) \right\}
  -d \left( l=\ln \left( \frac{r}{\al} \right) \right) \cos 2\theta_{r}
\right]
\virg 
\label{eqn:AA25}
\end{equation}
%------------------------------------------------------ 
  where other response functions are obtained in a similar way.
We note that the last term of Eq. (\ref{eqn:AA25}) is discarded 
  in Eqs. (\ref{eqn:Rsin-})$\sim$(\ref{eqn:Rcos+}), i.e., $R_A(r)$, 
  due to the condition $d \ll 1$ 
  in the present numerical calculation.
%---------------------------------------

%------------- Appendix B --------------
\section{Derivation of Eqs. (\protect\ref{eqn:K-app-1}) $\sim$ (\protect\ref{eqn:gm-app-1})}

 We examine the renormalization equations for 
   small $l/l_1$. 
   Noting that  
     $4t(l)/(\vf $ \\ $\al^{-1}) \ll 1$
  for the small $l/l_1$, we use the approximation that 
$J_0(4t(l)/(\vf \al^{-1})) \simeq 1 - \left\{ 4t(l)/(\vf \al^{-1}) \right\}^2/4$ 
and $J_1(4t(l)/(\vf \al^{-1})) \simeq \left\{ 4t(l)/(\vf \al^{-1}) \right\}/2$.
 By retaining  terms  up to the order of $(g_2 - g_2')^2$, 
 Eq. (\ref{eqn:rg-t}) is  calculated as 
%---------- (B1) ------------
\begin{equation}
\frac{4 t(l)}{\vf \al^{-1}} = 
{\tilde t} 
\exp \left[ \left( 1-\frac{g^2}{2} \right) l \right]
\virg
\label{eqn:A1}
\end{equation}
%---------------------------
where  ${\tilde t} \equiv 4t/(\vf \al^{-1})$ and 
$g \equiv g_2 - g_2'$. 
By rewriting $K_{\phi}(l)$  as  $K_\phi(l) = 1 + g_0(l)/2$ 
 and making use of Eq. (\ref{eqn:A1}),  
Eqs. (\ref{eqn:rg-K}), (\ref{eqn:rg-gp}) and (\ref{eqn:rg-gm}) 
  are calculated as 
%------------- (B2) ------------
\begin{eqnarray}
g_0(l) &=& \frac{{\tilde t}^2}{8} g^2 \left( \e^{2l} -1 \right)
\virg
\label{eqn:A2}\\
%------------------
g_{2\phi+}(l) &=& -g + \frac{{\tilde t}^2}{16} g^3 
\left( \e^{2l} -1 - 2l \right) \virg
\label{eqn:A3} \\
%------------------
g_{2\phi-}(l) &=& g + \frac{{\tilde t}^2}{16} g^3 
\left( \e^{2l} -1 - 2l \right) \virg
\label{eqn:A4} 
\end{eqnarray} 
%---------------------------------------
 which  correspond to  Eqs. (\ref{eqn:K-app-1}), 
(\ref{eqn:gp-app-1}) and (\ref{eqn:gm-app-1}), respectively.

%------------- Appendix C --------------
\section{Estimation of the Energy Gap}
 We examine Eqs. (\ref{eqn:rg-K})$\sim$(\ref{eqn:rg-t}) 
  with small coupling constants
  for  $l \gg l_1$, where  
the term including $J_\nu ( 4t(l)/(\vf \al^{-1}) )$ and 
  $g_{2 \phi +}$ can be disregarded. 
In this case, the renormalization group equations are rewritten as  
%---------- (C1), (C2) -------------
\begin{eqnarray}
\frac{d}{dl} K_\phi(l) &=& \frac{1}{2}g_{2\phi -}^2(l)
\virg
\label{eqn:C1} \\
\frac{d}{dl}  g_{2\phi -}(l) &=& \left( 2 - \frac{2}{K_\phi(l)} \right) 
g_{2\phi -}(l) \virg
\label{eqn:C2} 
\end{eqnarray}
%--------------------- 
where the initial condition is given by
$K_{\phi}(l_0)=1$ and $g_{2\phi-}(l_0)= g_2 -g_2'$ 
 with $l_0 = - \ln [4t/(\vf \alpha^{-1})]$. 
The differential equations (\ref{eqn:C1}) and (\ref{eqn:C2}) 
 can be integrated as 
%----------- (C3) ------------
\begin{eqnarray}
g_{2\phi -}^2(l) - 8 (K_\phi(l) - \ln K_\phi(l)) 
&=& g_{2\phi -}^2(l_0) - 8 (K_\phi(l_0) - \ln K_\phi(l_0)) \nonum \\
\simeq
g_{2\phi-}^2 (l) - g_0^2 (l) - 8 
&=& (g_2 - g_2')^2 - 8 \virg
\label{eqn:C3} 
\end{eqnarray} 
%----------------------------   
  where  $K_\phi(l) = 1 + g_0(l)/2$.
From Eqs. (\ref{eqn:C1})$\sim$(\ref{eqn:C3}),
one obtains  
%--------- (C6) ---------
\begin{equation}
g_0(l)=
|g_2-g_2'| \tan \left[ |g_2-g_2'| (l-l_0) \right]
\point
\label{eqn:C6}
\end{equation}
%------------------------
By defining the gap $\Delta$ as 
  $\Delta \equiv \vf \alpha^{-1} \exp [-l_{\Delta}]$ 
    with $g_{0}(l_{\Delta}) = 1$,  
  one obtains for $\Delta$,
%------------- (C8) ---------------
\begin{equation}
\Delta = 4t \exp \left[ -\frac{\pi}{2} \frac{1}{|g_2-g_2'|} +1 \right]
\point
\label{eqn:C8}
\end{equation}
%----------------------------------
From Eqs. (\ref{eqn:C3})$\sim$(\ref{eqn:C8})
the response function $R_{\sin \phi_-}(r) $ at 
  $l_\Delta(= \ln (r_\Delta /\al))$ 
  is calculated as
%------------- (C9) ------------------
\begin{eqnarray}
R_{\sin \phi_-}(r_{\Delta})
&=&
\exp \left[ 
 \int _0^{l_{\Delta}} dl 
 \left( -\frac{1}{K_\phi (l)} + g_{2\phi-}(l) \right)
\right]
\nonumber \\
&=& 
\left( \frac{4t}{\vf \al^{-1}} \right) ^{1-(g_2-g_2')}
\exp \left[ - \frac{\pi}{2} \frac{1}{|g_2-g_2'|} + 1 \right]
\nonumber \\
& & 
\times 
\left( 1+\frac{1}{(g_2-g_2')^2} \right)^{1/4}
\left( \frac{1}{|g_2-g_2'|} + \sqrt{ 1+\frac{1}{(g_2-g_2')^2} }\right)
,
\end{eqnarray}
%----------------------------------  
  which is shown as a function of $1/(g_2-g_2')$ in the inset in Fig. 4.

%-----------Reference--------
%==========================================


\begin{thebibliography}{99}
\bibitem{Yoshioka_err}
 There is an error of the the sign 
in the corresponding equation  in the previous calculation.
\cite{Yoshioka_Suzumura_JPSJ95}
\end{thebibliography}
\end{document}